\documentclass[%
aip,
amsmath,amssymb,
reprint,%
]{revtex4-1}

\usepackage{overpic}

\usepackage{siunitx}
\usepackage{graphicx}
\usepackage{bm}

\usepackage{etoolbox}

\makeatletter
\def\@email#1#2{%
 \endgroup
 \patchcmd{\titleblock@produce}
  {\frontmatter@RRAPformat}
  {\frontmatter@RRAPformat{\produce@RRAP{*#1\href{mailto:#2}{#2}}}\frontmatter@RRAPformat}
  {}{}
}%
\makeatother
\begin{document}

\preprint{AIP/123-QED}

\title{Circuit model for gate-induced excitation of quantum Hall edges tuned by DC gate bias}

\author{Kodai Sugizaki}
\affiliation{Department of Physics, Tohoku University, Sendai 980-8578, Japan}

\author{Takaaki Mano}
\affiliation{National Institute for Materials Science, Tsukuba, Ibaraki 305-0047, Japan}

\author{Go Yusa}%
\email{yusa@tohoku.ac.jp}
\affiliation{Department of Physics, Tohoku University, Sendai 980-8578, Japan}

\date{\today}

\begin{abstract}
  Edge excitations, which propagate along the edge of a quantum Hall system, are of great interest in the study of low-dimensional electron systems and mesoscopic transport phenomena.
  In this study, we investigate the relationship between the input voltage and the excitation waveform for gate-induced excitations.
  By varying the offset voltage applied to the gate, we observe that the excitation waveform changes from a first-derivative-like response to a response proportional to the input waveform in the time domain.
  In the frequency domain, we observe that the frequency dependence of the gate excitation changes significantly as the offset voltage crosses the pinch-off voltage.
  We show that these changes can be described by a single equivalent circuit model with different effective gate-2DEG capacitances on either side of the pinch-off voltage.
  In addition, we confirm that the effective capacitance obtained from the fitting of the model with the experimental data is consistent with the capacitance calculated from the device structure.
\end{abstract}

\maketitle

Quantum Hall edges, which exhibit chiral propagation and low-dissipation transport~\cite{QHEHalperin1982}, are promising platforms for quantum transport~\cite{Ji2003,nakamura2019aharonov} and quantum information processing~\cite{feve2008quantum,QETHallEffectYusa2011}.
One intriguing application is the simulation of an expanding universe.
Edge excitations propagate in one direction along the edge, so the edge can be regarded as a 1+1-dimensional spacetime in which the excitations play the role of light.
Expanding the edge then corresponds to the expansion of the 1+1-dimensional universe, and it has been theoretically proposed that it enables simulations of cosmological phenomena such as Hawking radiation~\cite{hotta2022expanding,nambu2023analog}.
For these applications, understanding and quantitatively modeling how edge excitations are generated is essential for controlling the excitations launched into the edge and interpreting their subsequent propagation.
Edge excitations can be generated either via  an ohmic contact ~\cite{EMPExperimentKamata2010,EMPExperimentgateungatedKumada2011},  or via a metal gate capacitively coupled to the edge~\cite{Matsuura2018,Kumada2020,InAs/InGaSbEMPKamata2022,kamiyama2022PRR,kamiyama2023APL,frigerio2024gate,Quentin2025,jeong2026spatiotemporal}.
Ohmic excitation produces an excitation signal proportional to the input voltage~\cite{EMPExperimentKamata2010,EMPExperimentgateungatedKumada2011}, because the input voltage directly modulates the electrochemical potential of the edge through the ohmic contact.
Although gate-induced edge excitation has been reported to produce a signal approximately proportional to the input voltage~\cite{Quentin2025,jeong2026spatiotemporal}, it can also exhibit a waveform resembling the first time derivative of the input voltage~\cite{kumada2013plasmon,Matsuura2018,InAs/InGaSbEMPKamata2022,kamiyama2022PRR,kamiyama2023APL}.
However, what determines whether gate excitation produces a waveform proportional to the input voltage or its time derivative remains unclear.
In this study, we investigate the offset-voltage dependence of gate-induced edge excitation through measurements in both the time and frequency domains.
Time-domain measurements reveal the response changes from derivative-like to proportional as the offset voltage is varied, while frequency-domain measurements systematically characterize its offset-voltage dependence.
We show that this transition can be explained by a single equivalent circuit model that accounts for the capacitive coupling between the gate and the two-dimensional electron gas in which the effective gate-2DEG capacitance differs between the two bias regimes and determines the characteristic frequency of the excitation response.

\begin{figure}
  \centering
  \includegraphics[width=\linewidth]{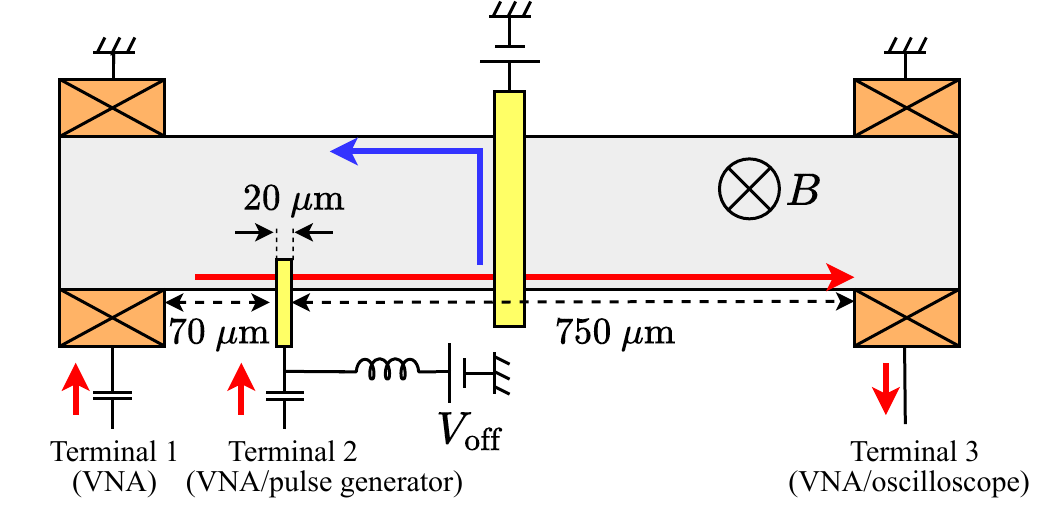}
  \caption{
    Schematic of the device structure and the frequency- and time-domain measurements setup.
    The orange and yellow elements indicate the ohmic contacts and metal gates, respectively.
    Edge excitations are generated by applying an AC voltage to either the ohmic contact (Terminal 1) or the metal gate (Terminal 2) and detected at the downstream ohmic contact (Terminal 3).
    The excitation gate has dimensions of $20~\mathrm{\mu m}$ along the edge and $50~\mathrm{\mu m}$ toward the bulk.
    Applying a negative voltage to the gate (center gate) at the center of the device depletes the 2DEG beneath it, blocking the propagation of edge excitations to the detecting ohmic contact.
    When exciting from either the ohmic contact or the  excitation gate, the other AC input port of the VNA is grounded.
  }
  \label{fig1}
\end{figure}

\begin{figure*}
  \centering
  \includegraphics[width=\linewidth]{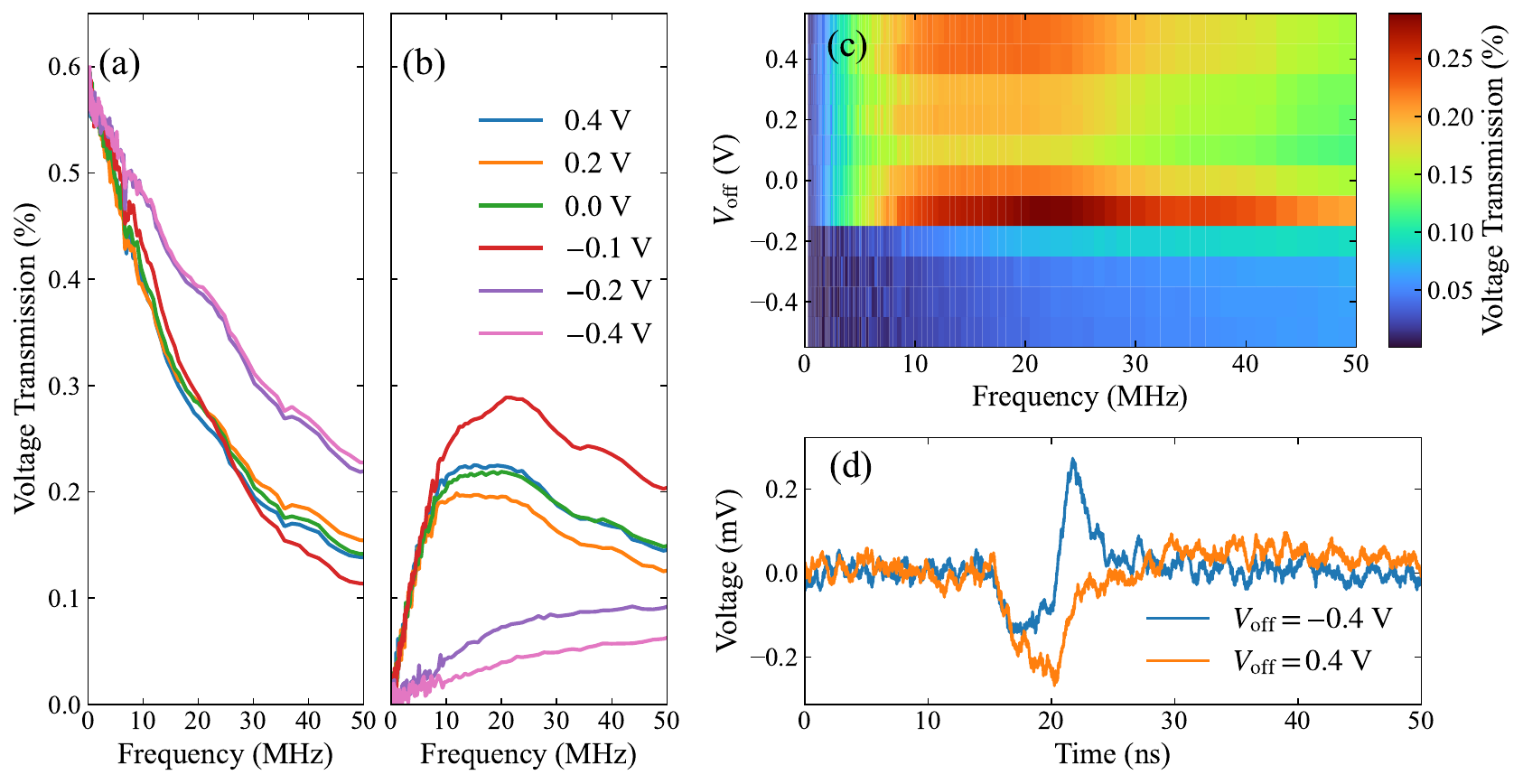}
  \caption{
    Time- and frequency-domain measurements of edge excitations at a magnetic field of $B=2$~$\mathrm{T}$.
    (a), (b) Frequency-domain measurements of voltage transmission by an ohmic contact and the metal gate, respectively.
    (c) Color map of the data in (b) as a function of frequency and $V_\mathrm{off}$.
    (d) Time-domain measurement of gate excitation.
  }
  \label{fig2}
\end{figure*}

We used GaAs/AlGaAs single heterostructures to form a two-dimensional electron gas (2DEG).
Fig.~\ref{fig1} shows a schematic of the device.
The electron density $n$ was $1.057\times 10^{11}~\mathrm{cm^{-2}}$.
Ni/AuGe/Ni and Al were used for the ohmic contacts and gate electrodes, respectively.
The pinch-off voltage of the gates is approximately $V_{\text{pinch-off}}=-0.15~\mathrm{V}$.

Edge excitations were generated by applying an AC signal to either the ohmic contact or the gate (excitation gate).
For gate excitation, a DC offset voltage $V_{\mathrm{off}}$ and an AC signal were simultaneously applied to the gate through a bias tee. The resulting edge excitations were detected at a downstream ohmic contact.

In the frequency-domain measurements, the AC signal was generated by a vector network analyzer (VNA).

The input power was set to $-24$~dBm, corresponding to a voltage amplitude of 20~mV for a 50~$\Omega$  load and 40 mV under open-circuit conditions.
These measurements were conducted at temperatures of 20--40~mK.
In the time-domain measurements, a pulse voltage was applied to the excitation gate using a pulse generator.
The pulse rise time, amplitude, period, and width were $90~\mathrm{ps}$, $100~\mathrm{mV}$, $50~\mathrm{ns}$, and $5~\mathrm{ns}$, respectively. The signal was amplified using a room-temperature amplifier (gain: $18.8~\mathrm{dB}$, frequency range: 20~MHz--3~GHz) and detected with an oscilloscope.
Time-domain measurements were conducted at a temperature of $1.5~\mathrm{K}$.

Fig.~\ref{fig2}(a) shows the frequency dependence of the transmission for ohmic excitation.
The measured signal contains crosstalk generated by direct capacitive coupling in the measurement setup, such as the printed circuit board (PCB) and bonding wires.
The data shown in Fig.~\ref{fig2}(a)--(c) were obtained by subtracting the crosstalk-only signal, which was measured at negative magnetic fields, where the edge excitation does not reach the detection ohmic contact due to the chirality.

After subtracting the crosstalk-only signal, the transmission for ohmic excitation decreases with increasing frequency.
This decrease is attributed to attenuation in the measurement system, including the wiring (cables, PCB, bonding wires, etc.), as well as dissipation during propagation along the edge.
It is known that edge excitations can be attenuated during propagation due to dissipation associated with nonzero $R_{xx}$ and coupling to charge puddles, which are localized regions of electron accumulation caused by disorder-induced potential fluctuations in the bulk of the 2DEG~\cite{Johnson2003,frigerio2024gate}. Through this coupling, energy is transferred from the edge excitations to the charge puddles, resulting in attenuation of the edge excitations.
Notably, the transmission for ohmic excitation depends on the offset voltage applied to the excitation gate, even though the gate is not used to generate the excitation. The dependence is most pronounced across the pinch-off voltage $V_{\text{pinch-off}}$: the transmission for $V_{\mathrm{off}}> V_{\text{pinch-off}}(=-0.15~\mathrm{V})$ is smaller than that for $V_{\mathrm{off}} < V_{\text{pinch-off}}$, while variations within each regime are comparatively small.
This reduction is attributed to attenuation associated with screening by the excitation gate~\cite{EMPExperimentgateungatedKumada2011,Kumada2020}.
When $V_{\mathrm{off}}>V_{\text{pinch-off}}$, edge excitation propagates beneath the excitation gate and its electric field is capacitively coupled to the gate, giving rise to an ac displacement current that can be a source of attenuation.

In contrast, gate excitation exhibits a markedly different frequency dependence compared to ohmic excitation [Fig.~\ref{fig2}(b) and (c)].
For $V_{\mathrm{off}}<V_{\text{pinch-off}}$, the voltage transmission generally increases with increasing frequency, at least below $50~\mathrm{MHz}$.
For $V_{\mathrm{off}}>V_{\text{pinch-off}}$, it also increases with increasing frequency below approximately $20~\mathrm{MHz}$, but decreases monotonically at higher frequencies.

Given the distinct frequency dependencies on either side of $V_{\text{pinch-off}}$, we examined the pulse waveforms at $V_{\mathrm{off}}=-0.4$ and $+0.4~\mathrm{V}$ in the time domain [Fig.~\ref{fig2}(d)].
In this measurement, the crosstalk-only signal was obtained by applying a negative voltage to the center gate to block the edge excitation and was subtracted from the original signal.
At $V_{\mathrm{off}}=+0.4~\mathrm{V}$, the signal shows a pulse-like response, whereas at $V_{\mathrm{off}}=-0.4~\mathrm{V}$, it resembles the first derivative of the input pulse.


\begin{figure*}
  \centering
  \begin{overpic}[width=\linewidth]{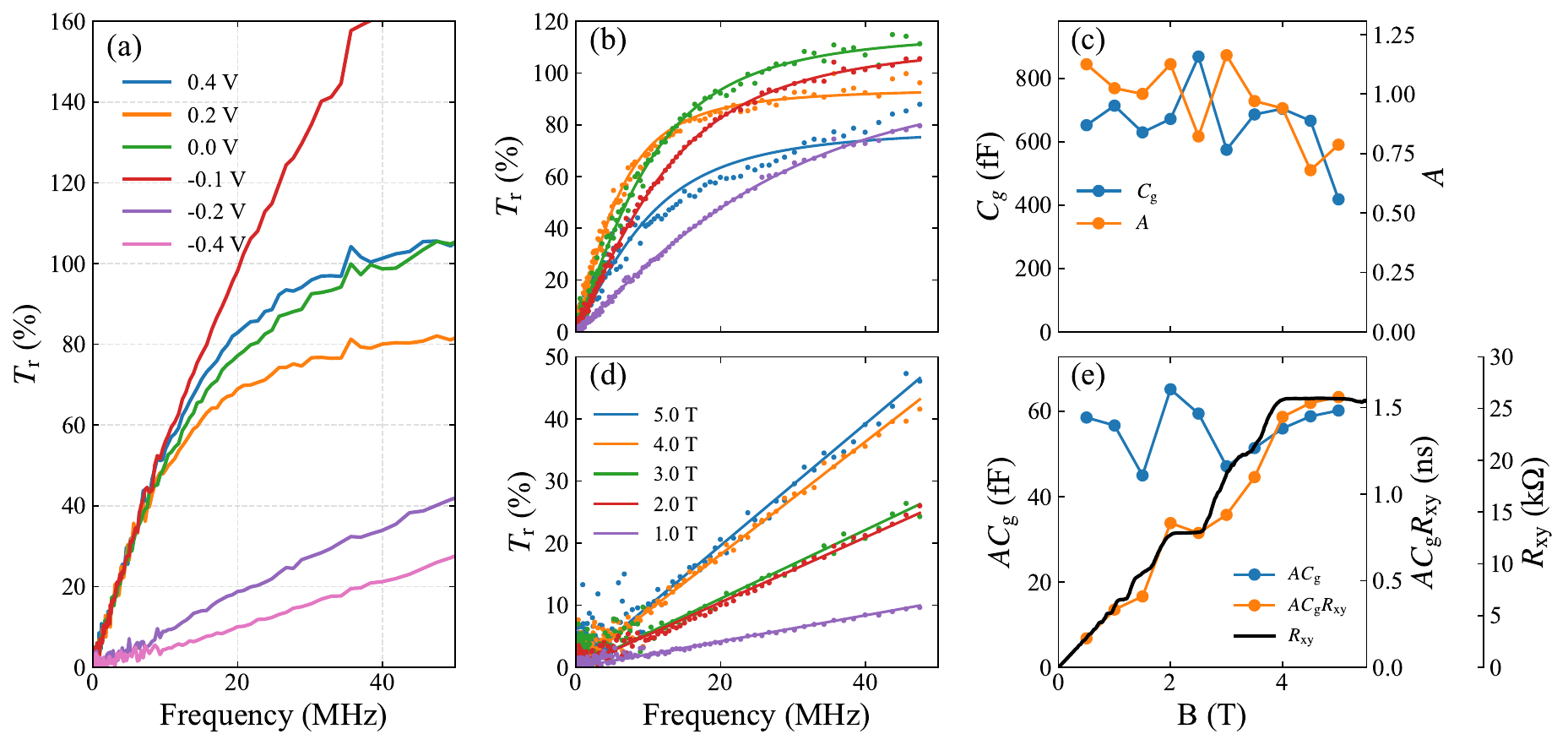}
    \put(69,26.5){\includegraphics[width=0.17\linewidth]{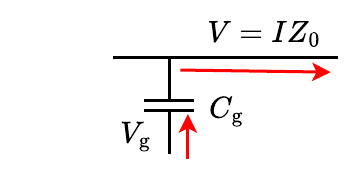}}
  \end{overpic}
  \caption{
  (a) Frequency dependence of $T_\mathrm{r}$ at various values of $V_{\mathrm{off}}$ at $B=2~\mathrm{T}$.
  (b),(d) Frequency dependence of $T_\mathrm{r}$ and fitting curves at various $B$ for $V_{\mathrm{off}}=+0.4~\mathrm{V}$ and $-0.4~\mathrm{V}$, respectively.
  Markers represent experimental data, and solid lines represent fitting curves.
  (c) Fitting parameters at various $B$ for $V_{\mathrm{off}}=+0.4~\mathrm{V}$.
  $C_{\mathrm{g}}$ (blue) and $A$ (orange).
  (e) Fitting parameters for $V_{\mathrm{off}}=-0.4~\mathrm{V}$ and Hall resistance $R_{xy}$ at various $B$.
  $AC_{\mathrm{g}}$ (blue), $AC_{\mathrm{g}}R_{xy}$ (orange) and $R_{xy}$ (black).
  }
  \label{fig3}
\end{figure*}

To understand these characteristic responses of gate excitation, we distinguish the contribution of the excitation process from those of edge propagation and the measurement system. The measured voltage transmission $T$ can be expressed as
$T = T_{\mathrm{ex}}T_{\mathrm{prop}}T_{\mathrm{wiring}}.$
Here, $T_{\mathrm{ex}}$ is the voltage transmission from the ohmic contact or gate to the edge, $T_{\mathrm{prop}}$ is the transmission along the edge from excitation to detection, and $T_{\mathrm{wiring}}$ is the transmission associated with the wiring and measurement system.
The distance from the ohmic contact to the excitation gate is approximately $70~\mathrm{\mu m}$, which is much shorter than the $750~\mathrm{\mu m}$ distance from the excitation gate to the detecting ohmic contact. We therefore approximate the ratio of $T_\mathrm{prop}$ for ohmic and gate excitation as
$\frac{T_{\mathrm{gate,prop}}}{T_{\mathrm{ohmic,prop}}}\approx1$,
allowing for a finite error in this approximation.
Also, we assume that $T_{\mathrm{wiring}}$ is the same for ohmic and gate excitation.

To estimate the magnitude of the propagation error, we assume an exponential dependence of the propagation transmission on frequency and distance~\cite{Johnson2003,frigerio2024gate}, such that
$T_{\mathrm{prop}}\propto\exp[-k(f)x]$,
where $k(f)$ is the frequency-dependent attenuation coefficient and $x$ is the propagation distance. In addition, we assume $k(0)\approx0$, consistent with the negligible longitudinal resistance $R_{xx}\ll R_{xy}$ in the quantum Hall regime.

Under this assumption, the ratio of $T_\mathrm{prop}$ for gate and ohmic excitation at $f=f_0$ is given by
\begin{equation}
  \frac{T_{\mathrm{gate,prop}}}{T_{\mathrm{ohmic,prop}}}
  =
  \left[
    \frac{T_{\mathrm{ohmic,prop}}(f_0)}
    {T_{\mathrm{ohmic,prop}}(f=0)}
    \right]^{
  \frac{x_{\mathrm{gate}}-x_{\mathrm{ohmic}}}{x_{\mathrm{ohmic}}}},
\end{equation}
where $x_\mathrm{gate}=750~\mathrm{\mu m}$ is the distance from the excitation gate to the detecting ohmic contact and $x_{\mathrm{ohmic}}={840}~\mathrm{\mu m}(=750+20+70)$ is the distance from the excitation ohmic contact to the detecting ohmic contact.
To conservatively estimate this error, we attribute the entire frequency-dependent attenuation observed for ohmic excitation [Fig.~\ref{fig2}(a)] to propagation along the edge, without subtracting attenuation associated with the wiring and measurement system.
Here, we take the transmission at $f={300}~\mathrm{kHz}$ as the reference value for the low-frequency response.
Under these assumptions, the deviation of the propagation-transmission ratio from unity is up to approximately $31.2~\mathrm{\%}$ at ${50}~\mathrm{MHz}$ for $V_{\mathrm{off}}={-0.1}~\mathrm{V} , B={5}~\mathrm{T}$.

To isolate the frequency dependence associated with the gate excitation process from that of edge propagation and the measurement system, we take the ratio of the voltage transmissions for gate and ohmic excitation.
\begin{equation}
  T_{\mathrm{r}}
  =
  \frac{T_{\mathrm{gate}}}{T_{\mathrm{ohmic}}}
  \approx
  \frac{T_{\mathrm{ex,gate}}}{T_{\mathrm{ex,ohmic}}}.
\end{equation}
Here, $T_{\mathrm{gate}}$ and $T_{\mathrm{ohmic}}$ are the voltage transmissions for gate and ohmic excitation, respectively, while $T_{\mathrm{ex,gate}}$ and $T_{\mathrm{ex,ohmic}}$ denote the voltage transmissions from the excitation gate and the ohmic contact to the 2DEG, respectively.
Since the excitation gate screens both the edge excitations generated by the ohmic contact and those generated by the gate itself, $T_{\mathrm{r}}$ is calculated using $T_{\mathrm{ohmic}}$ and $T_{\mathrm{gate}}$ measured at the same $V_{\mathrm{off}}$.

The resulting $T_{\mathrm{r}}$ is approximately proportional to frequency for $V_{\mathrm{off}}<V_{\text{pinch-off}}$, whereas it tends to saturate at high frequencies for $V_{\mathrm{off}}>V_{\text{pinch-off}}$  [Fig.~\ref{fig3}(a)].
This difference in frequency dependence can be explained using an equivalent circuit model based on the chiral distributed-element (CDE) model\cite{hashisaka2013distributed}, in which the quantum Hall edge is represented as a chiral transmission line with an impedance $Z_0=R_{xy}$, where $R_{xy}$ is the Hall resistance [Fig.~\ref{fig3}(c) inset].
Here, $V(x,t)$ is the voltage of the edge, and $I(x,t)$ is the current flowing along the edge.
Let $C_g$ be the effective capacitance between the excitation gate and the 2DEG, and let $V_g$ be the gate voltage.
The current flowing from the gate to the edge is then given by $I=C_g\frac{d}{dt}(V_g-V)$.
Let $V_g=V_\mathrm{ac}+V_\mathrm{off}$ and $V_\mathrm{ac}\propto\exp(i\omega t)$.
The voltage transmission from the excitation gate to the edge $T_{\mathrm{ex,gate}}$, is given by
\begin{equation}
  T_{\mathrm{ex,gate}}
  =
  \frac{i\omega C_g Z_0}
  {1+i\omega C_g Z_0}.
  \label{gate_transmittance}
\end{equation}
Here, we assume that $C_g$ is constant.

From this equation, when $\omega C_g Z_0 \ll 1$, $T_{\mathrm{ex,gate}} \approx i \omega C_g Z_0$.
Thus, the voltage transmission is proportional to frequency, and the waveform corresponds to the first derivative of the input waveform.
In contrast, when $\omega C_g Z_0 \gg 1$, $T_{\mathrm{ex,gate}}\approx 1$,
so that the voltage transmission becomes independent of frequency and the waveform becomes proportional to the input waveform.

We assume that \(T_{\mathrm{ex,ohmic}}\) is frequency independent because the input voltage directly modulates the electrochemical potential of the edge through the ohmic contact.
Using Eq.~\eqref{gate_transmittance}, we introduce the difference in excitation efficiency between gate and ohmic excitation as a frequency-independent fitting parameter, $A>0$.
$T_{\mathrm{r}}$, is then expressed as

\begin{equation}
  T_{\mathrm{r}}
  =
  A
  \left|
  \frac{i\omega C_g Z_0}
  {1+i\omega C_g Z_0}
  \right|
  \label{fitting_model}
\end{equation}

Fig.~\ref{fig3}(b) and (c) show $T_{\mathrm{r}}$ and the fitted parameters $A$ and $C_g$ at $V_{\mathrm{off}}>V_{\text{pinch-off}}$($V_{\mathrm{off}}={+0.4}~\mathrm{V}$), respectively.
Fitting the data using Eq.~\eqref{fitting_model} yields $A=0.96 \pm 0.15$ and $C_g=(658 \pm 108)~\mathrm{fF}$, where the uncertainties represent standard deviations.
Here, we use measured Hall resistance $Z_0=R_{xy}$ [Fig.~\ref{fig3}(e)].

For comparison, the capacitance $C_g$ assuming a parallel-plate capacitor between the gate and the 2DEG,
$C_g=\varepsilon_r\varepsilon_0\frac{S}{d}\approx {732}~\mathrm{fF}$, is comparable to the fitted value.
Here, $\varepsilon_r=12.4$ is the relative permittivity of GaAs at 40~mK~\cite{Madelung1996},
$S={20}~\mathrm{\mu m}\times{50}~\mathrm{\mu m}$ is the gate area,
and $d={150}~\mathrm{nm}$ is the distance between the gate and the 2DEG.

In contrast,  at $V_{\mathrm{off}}<V_{\text{pinch-off}}$ ($V_{\mathrm{off}}=-0.4~\mathrm{V}$), $T_{\mathrm{r}}$ is approximately proportional to frequency [Fig.~\ref{fig3}(d)]. This behavior is consistent with $T_{\mathrm{r}}=A\omega C_g Z_0$ in the limit of $\omega C_g Z_0 \ll 1$.

In this regime, $A$ and $C_g$ cannot be determined independently, and only the slope $2\pi AC_gZ_0$ of $T_{\mathrm{r}}$ as a function of frequency is obtained directly from the fit.
The slope of $T_\mathrm{r}$ increases with increasing magnetic field and exhibits plateau-like structures around $B\approx$2--3~T and 4--5~T, corresponding approximately to the $\nu=2$ and $\nu=1$ quantum Hall regimes, respectively [Fig.~\ref{fig3}(e)].
Divided by $Z_0=R_{xy}$, we obtain $AC_g=(55.8\pm 5.9)$~fF.

Assuming that $A$ is the same as that at $V_{\mathrm{off}}={+0.4}~\mathrm{V}$, the effective capacitance is estimated to be $C_g\approx{58.1}~\mathrm{fF}$, about an order of magnitude smaller than that at $V_{\mathrm{off}}={+0.4}~\mathrm{V}$.

The difference in $C_g$ between $V_\mathrm{off}={+0.4}~\mathrm{V}$ and $V_\mathrm{off}={-0.4}~\mathrm{V}$ is considered to be due to whether the 2DEG beneath the gate is depleted.
For $V_{\mathrm{off}}<V_{\text{pinch-off}}$, the 2DEG beneath the gate is depleted, and the gate can couple capacitively to the 2DEG only in a limited region near the gate edge.
In contrast, for $V_{\mathrm{off}}>V_{\text{pinch-off}}$, the 2DEG remains beneath the gate, allowing capacitive coupling to the 2DEG beneath the entire gate area, resulting in a larger effective capacitance.
This difference of $C_g$ explains the change in the frequency dependence as $V_{\mathrm{off}}$ crosses the pinch-off voltage [Fig.~\ref{fig3}(a)] and the transition between first-derivative-like and proportional responses observed in the time-domain measurements [Fig.~\ref{fig2}(d)].

On the other hand, at $V_{\mathrm{off}}=-0.1~\mathrm{V}\approx V_{\text{pinch-off}}(={-0.15}~\mathrm{V})$, the frequency dependence of $T_\mathrm{r}$ is significantly different from that at $V_{\mathrm{off}}=-0.4~\mathrm{V}$ and $+0.4~\mathrm{V}$ [Fig.~\ref{fig3}(a)].
This behavior is considered to be because $C_{\mathrm{g}}$ changes significantly with the excitation voltage applied to the gate.
In this regime, the constant-capacitance model [Eq.~\eqref{gate_transmittance}] is not appropriate for describing the measured frequency dependence.

In this study, we investigated the excitation characteristics of quantum Hall edges induced by a gate.
We observed that the excitation waveform changes from a derivative-like response to a proportional response to the input waveform as the offset voltage $V_{\mathrm{off}}$ is varied.
This behavior is described by a single equivalent circuit model of the edge, in which the characteristic response is governed by $C_g\omega Z_0$.
Furthermore, the value of $C_g$ obtained at $V_{\mathrm{off}}={+0.4}~\mathrm{V}$ is consistent with the capacitance calculated by modeling the gate and 2DEG as a parallel-plate capacitor.

\begin{acknowledgments}
  The authors thank T. Takayanagi, M. Hotta, J. N. Moore, and Y. Jeong for fruitful discussions. This work was supported by JSPS KAKENHI Grant Numbers~JP21H05182, JP21H05188, and JP24H00399. We are grateful to the YITP workshop YITP-T-25-01 held at YITP, Kyoto University, where part of this work was carried out.
\end{acknowledgments}

\section*{Data Availability Statement}

The data that support the findings of this study are available from the corresponding author upon reasonable request. 

\bibliographystyle{aipnum4-2}
\bibliography{references}

\end{document}